\documentclass[summary]{ursi}

\DeclareMathOperator*{\argmin}{arg\,min}
\newcommand{\norm}[1]{\left\lVert#1\right\rVert}

\usepackage{amsmath}

\title{A Compressed Sensing Faraday Depth Reconstruction Framework for the MeerKAT MIGHTEE-POL Survey.}

\author{Miguel C{\'a}rcamo*\affref{ref1}\affref{ref7}\affref{ref8}, Anna Scaife\affref{ref1}\affref{ref2}, Russ Taylor\affref{ref3}, Matt Jarvis\affref{ref4}\affref{ref9}, Micah Bowles\affref{ref1}, Srikrishna Sekhar\affref{ref3}\affref{ref5}, Lennart Heino\affref{ref3}  and Jeroen Stil\affref{ref6}}

\affiliation{%
  \aff{ref1}{Jodrell Bank Centre for Astrophysics, Department of Physics and Astronomy, University of Manchester, Manchester, UK}
  \aff{ref2}{The Alan Turing Institute, Euston Road, London, NW1 2DB, UK}
  \aff{ref3}{Inter-Unversity Institute for Data Intensive Astronomy, Cape Town, 
South Africa}
  \aff{ref4}{Astrophysics, Department of Physics, University of Oxford, Keble Road, Oxford, OX1 3RH, UK}
  \aff{ref5}{National Radio Astronomy Observatory, 1003 Lopezville Road, Socorro, NM 87801, USA}
  \aff{ref6}{Department of Physics and Astronomy, University of Calgary, 2500 University Drive NW, Calgary AB T2N 1N4, Canada}
  \aff{ref7}{University of Santiago of Chile (USACH), Faculty of Engineering, Computer Engineering Department, Chile}
  \aff{ref8}{Center for Interdisciplinary Research in Astrophysics and Space Exploration (CIRAS), Universidad de Santiago de Chile}
  \aff{ref9}{Physics Department, University of the Western Cape, Private Bag X17, Bellville, 7535, South Africa}
}

\begin{document}

\maketitle

\begin{abstract}
  In this work we present a novel compute framework for reconstructing Faraday depth signals from noisy and incomplete spectro-polarimetric radio datasets. This framework is based on a compressed-sensing approach that addresses a number of outstanding issues in Faraday depth reconstruction in a systematic and scaleable manner. We apply this framework to early-release data from the MeerKAT MIGHTEE polarisation survey. 
\end{abstract}

\section{Introduction}

The MeerKAT MIGHTEE survey is imaging four extragalactic fields, COSMOS, XMM-LSS, CDFS, and ELAIS\,S1 \cite{Jarvis:20184F}.  All fields are being observed at L-band from 880--1680 MHz, in multiple pointings that will be mosaiced to a final image with a broad-band sensitivity of $\sim$2\,$\mu$Jy\,beam$^{-1}$ \cite{mightee}.  The MIGHTEE project is creating science data products for total intensity, broad-band continuum science, HI spectra-line science, and spectro-polarimetric science.  
Initial data sets for shared-risk early science projects are being produced in each category from early observations of COSMOS and XMMLSS. 

Next-Generation radio telescopes such as the Square Kilometre Array (SKA) will produce raw data volumes orders of magnitude larger than those from current facilities. Already the data volumes from existing SKA precursors such as MeerKAT are revealing big data and high-throughput computing (HTC) challenges. An example of this is the study of cosmic magnetism using Faraday Rotation information. In this scenario, multiple lines of sight from a linearly polarized spectral image cube must be passed through a non-linear optimization algorithm in order to obtain a 3D Faraday depth reconstruction. Addressing this problem not only involves the application of parallelism and HTC techniques, but also the introduction of software paradigms that facilitate automated workflows and pipeline operations.

This paper describes the application of a novel Faraday compute framework for compressed sensing (CS) reconstruction of spectro-polarimetric data from the MIGHTEE polarisation (MIGHTEE-POL) science case.

\section{CS for Faraday depth reconstruction}

Compressed sensing (CS) is a novel paradigm that contradicts the Nyquist-Shannon theorem for data acquisition under certain circumstances. The degree of success of this method depends on the amount of information that can be supplied to constrain the signal solution while being consistent with the measurements \cite{thompsonbook}. These constraints are sparsity, non-negativity, compactness, and the smoothness of the signal. To fully recover a signal, $x$, a compressible representation as a $k$-sparse vector must be imposed. Mathematically, a signal is defined to be sparse or compressible if it contains a small number, $k$, of non-zero or significant values. Typically, signals are not themselves sparse, but they can have a sparse representation in some basis $\Phi$. In this case we can refer to $x$ being $k$-sparse considering $x=\Phi c$, where $\norm{c}_0 \leq k$. The main problem with this approach is that the zero-norm function, which counts the non-zero elements of a vector, is non-convex and finding a solution that approximates the true minimum is generally a non-deterministic polynomial-time hard (NP-hard) problem and computationally very intensive \cite{np-hard}.

However, \cite{donoho_cs} and \cite{candestao2016} discovered that under general conditions it is possible to solve (relax) the above problem by using basis pursuit or enforcing convexity in the $\norm{\cdot}_0$. That is, considering a Laplacian prior or a L1 regularization as
\begin{equation}
\hat{x} = \argmin_{x} \norm{x}_1 \; \text{subject to} \; x\in \mathcal{B}(y),
\label{eq:l1min}
\end{equation}
and that $\mathcal{B}(y)$ is convex, then the problem has become computationally feasible and can be solved with efficient methods from convex optimization. Here, the $L_1$ norm of a vector is defined as the sum of the absolute values of each component of the vector.
%


In the presence of noise, Equation~\ref{eq:l1min} can also be written as an unconstrained problem:
\begin{equation}
\hat{x} = \argmin_{x} \frac{1}{2}\norm{Ax-y}_2^2 + \eta \norm{x}_1\; ,
\label{eq:l1min_2}
\end{equation}
where the first term is equivalent to the squared sum of the data residuals and $\eta$ is a regularization parameter that determines the relative importance between minimizing the $L_1$-norm and the measurement residuals.

For reconstruction of Faraday depth signals, one of the first works in this area was \cite{frick}, which demonstrated the applicability of the known Mexican Hat wavelet to decompose the Faraday depth Spectrum. Later \cite{li20102} reconstructed synthetic data from \cite{FRM} and used three algorithms for the reconstruction of different kinds of structure. The difference between the algorithms lies in which space these structures are sparse. Another work that reconstructs Faraday depth spectrum signals using CS was \cite{Andrecut_2011}. This work uses the Matching-Pursuit (MP) algorithm, an L1-norm regularization and a boxcar dictionary in order to reconstruct both thin and thick Faraday signals. The work of \cite{akiyama2018faraday} used a trade-off between sparsity and smoothness in Faraday depth space using regularizations such as the $L_1$-norm and Total Variation (TV) or Total Squared Variation (TSV). This used a Faraday depth model from \cite{Ideguchi_2014} with both Faraday thin and thick structures in a frequency band from $300$ to $3000$\,MHz to test the procedure. The results were compared with an RMCLEAN version from \cite{10.1093/pasj/psw039}. The work made by \cite{Cooray_2020} improved on this last using an iterative restoration algorithm based on the projected gradient descent where at each iteration, different assumptions of the Faraday depth spectrum can be made such as sparsity and considering that some parts of the spectrum are the result of the Rotation Measure spread function (RMSF). In this work the smoothness of the polarization angle can also be added as a constraint.

There are two things to note across these two most recent works. Both create data in Faraday depth space and then transform into $\lambda^2$-space using a DFT. However, doing this implies that the $\lambda^2$-space is implicitly regularly-spaced. It is worth noting that an irregular $\lambda^2$-space not only adds an additional layer of noise to the problem but also a problem similar to that seen in image synthesis. Secondly the simulations in these works do not account for frequency channel excision due to RFI, instrumental problems or calibration.

\section{The {\tt cs-romer} framework}

In this work we propose a CS-framework that solves the problem statement in Equation~\ref{eq:l1min_2}, extending the objective function to impose edge-smoothed constraints when needed, such that
\begin{equation}
    \hat{x} = \argmin_{x} \Biggl\{ f(x) + g(x) \Biggr\} ,
    \label{eq:gen-objf}
\end{equation}
%
%
%
where $f(x)$ is commonly seen in the literature as the data or $\chi^2$ term, which is a function of the observed polarized intensity measurements as a function of $\lambda^2$, the model measurements or signal estimate, and the variance for each channel. $\chi^2$ is considered a convex smooth function and $g(x)$ is a convex function that is non-differentiable in some region such that
\begin{equation}
\nonumber    g(x) = \eta_1 \text{L1}(x) + \eta_2 {\rm TV^{\ast}}(x)\;,
    \label{eq:objf}
\end{equation}
where $x$ can be the signal itself or a wavelet representation and TV$^{\ast}$ is either the TV or TSV norm. 

First, the framework calculates $\lambda^2_{\text{min}}$ and $\lambda^2_{\text{max}}$. Then, $\Delta \lambda^2$ and $\delta \lambda^2$ are computed, defining the maximum observable Faraday depth, the resolution in Faraday depth space and the largest scale in Faraday space to which the data are inherently sensitive. The framework selects the cell-size in Faraday depth space as $\phi_R = \delta \phi / \rho$, where $\rho$ is the oversampling factor which can be selected as $4$ or $5$. We can use $||\phi_{\text{max}}||$ to calculate the length of the grid in Faraday depth space. Since our observed measurements belong to an irregular space and Faraday depth space is on a regular grid the framework offers two options to estimate the model data. The most intuitive option is perhaps to grid the observed measurements, in which case the framework uses a DFT or FFT.  The second option is to use the Non-Uniform Fast Fourier Transform (NUFFT), which performs an FFT on non-uniform sampled data using an FFT on the signal and then a min-max interpolation. 

Having decided on a method with which to approximate the model data points, the next step is to use an optimization method to minimize the objective function (see Equation \ref{eq:gen-objf}). In this work we have chosen the Fast Iterative Shrinkage-Thresholding Algorithm (FISTA) \cite{fista}. This method is an enhancement of the ISTA method and has been used before in Faraday spectra reconstruction \cite{li20102}. 

Many attempts have been made to automatically calculate the regularization parameters, $\eta_1$ and $\eta_2$, e.g. \cite{KARL2005183, Hansen00thel-curve, Belge_2002, SHI201872}. For this framework, and only for L1 regularization, we adopt the error bound calculation in \cite{2014MNRAS.439.3591C, purify2}. Additionally, we have adopted the adaptive regularization method from \cite{li20102} where on each iteration $1\sigma$ is subtracted from $\eta_1$.

In the case where the signal can be represented by a wavelet dictionary, our framework offers the set of discrete wavelet transforms (DWT) from the package \texttt{pywt} \cite{Lee2019}. However, sometimes the DWT can cause problems due to its shift variance and poor directional properties \cite{uwt-denoising}. As an alternative, and as a way to reconstruct both thin and thick Faraday structures, we have added the Undecimated Wavelet Transform (UWT) 1D functions from \cite{Lee2019} to our framework. 
%
\begin{figure*}[h!]
    \centering
    \includegraphics[width=0.9\textwidth]{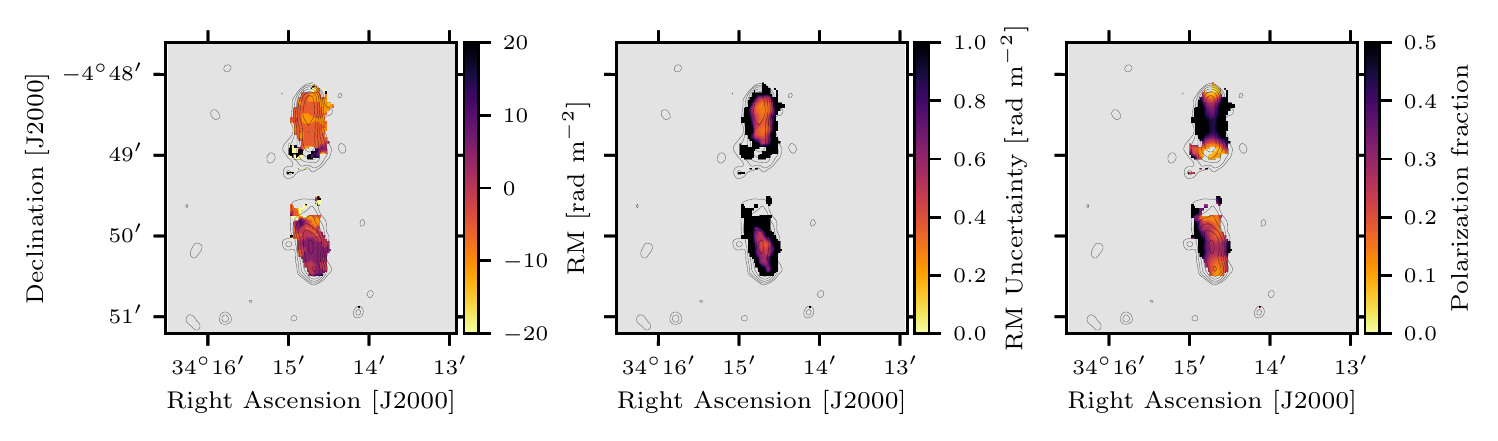}
    \caption{XMM-LSS Radio galaxy RM. From left to right: the rotation measure, the rotation measure uncertainty at the peak and the polarization fraction. The contours show the total intensity image at 5\,$\sigma$ with increments of 2\,$\sigma$.}
    \label{fig:radiogal-test}
\end{figure*}

\begin{figure*}[h!]
    \centering
    \includegraphics[width=0.9\textwidth, height=0.17\textheight]{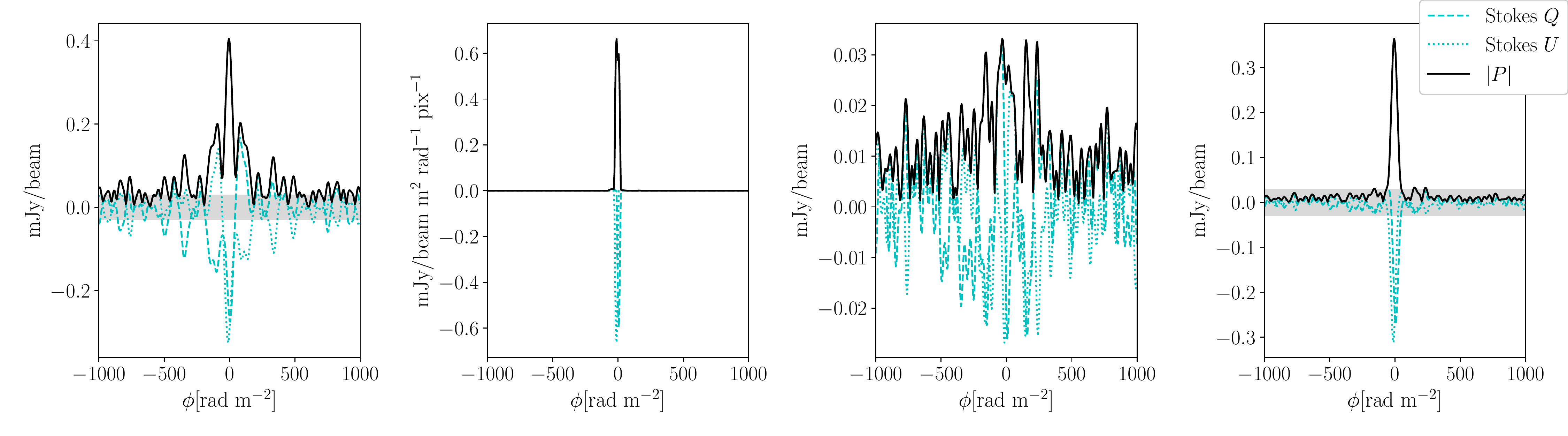}
    \caption{Single LOS from the radio galaxy in Figure~\ref{fig:radiogal-test}, see text for details. From left to right: the dirty, the model, the residual and the restored Faraday spectra; $\pm$5\,$\sigma_{\phi}$ limits are shown as a shaded area.}
    \label{fig:radiogal-los-test}
\end{figure*}

\section{Wavelet and performance evaluation}

Depending on the data, Faraday depth signals might have different structures. i.e thin sources, thick sources or both. To select the wavelet which best represents the signal, we simulate data given a known radio-telescope configuration with {\tt cs-romer}. We use these data to create multiple realisations of data flagging and noise levels, which are then reconstructed using the full set of wavelet families. From these we select the best wavelet by evaluating the peak signal-to-noise ratio (PSNR), root mean squared error (RMSE), the Akaike's Information Criterion (AIC) and Bayesian Information Criterion (BIC). In other words, we select the wavelet which results in the reconstruction with highest PSNR, lowest RMSE and lowest AIC and BIC.

\section{Application to MIGHTEE-POL data}

MIGHTEE-POL QU floating-point data cubes have dimensions of 6144$\times$6144$\times$239$\times$2, making them nearly 67.22\,GB in size. Moreover, the regular-spaced Faraday Depth cubes are 6144$\times$6144$\times$373$\times$2 for XMM-LSS, which results in a data volume of 104.9\,GB. Therefore the reconstruction of one field from the MIGHTEE-POL survey data requires $\sim$160\,GB of memory. Even though this can be seen as a big data problem, each line of sight (LOS) can be processed independently and therefore, a big computing problem arises since it is not optimal to run 6144$\times$6144 optimizations sequentially. Thus, to speed up the reconstruction we use the {\tt joblib} python library. Moreover, we only make calculations on those lines of sight where total intensity is over 3$\sigma_I$ and polarization intensity over 3$\sigma_P$. This decreases the total number of LOS by 99.75\% for the XMM-LSS field. Here we use a canonical spectral index value of 0.7 for all LOS, but more generally the {\tt cs-romer} framework can accept a {\tt casa} spectral index map. The Faraday depth parameters for the XMM-LSS field are $\delta \phi = 41.6$, max-scale$=97.3$ and $||\phi_{\text{max}}|| = 2057.7$, all in units of rad\,m$^{-2}$. The reconstruction takes 81.7 minutes on a server with 2 Intel(R) Xeon(R) Gold 5222 3.80\,GHz CPUs and 1.5\,TB RAM. Figure~\ref{fig:radiogal-test} shows the reconstructed RM, the RM uncertainty and the polarization for a radio galaxy selected from the XMM-LSS survey field. Additionally, Figure~\ref{fig:radiogal-los-test} shows the dirty, model, residual and restored Faraday spectra for the LOS with highest Stokes~I intensity from this radio galaxy.


\section{Conclusions}
\label{sec:conclusion}

In this work we have presented the {\tt cs-romer} compute framework, a CS approach to reconstructing Faraday depth spectra from noisy and incomplete spectro-polarimetric radio measurements. We have demonstrated that {\tt cs-romer} can be used in a big-data and HTC scenario such as that appropriate for MIGHTEE-POL. In future work, we intend to decrease the reconstruction time further by distributing the framework across multiple compute nodes and exploring the impact of big data libraries such as {\tt dask}.

\section{Acknowledgements}

MC acknowledges support from ANID PFCHA/DOCTORADO BECAS CHILE/2018-72190574. AMS and MB gratefully acknowledge support from the UK Alan Turing Institute under grant reference EP/V030302/1. MB gratefully acknowledges support from the UK Science \& Technology Facilities Council (STFC).
\normalsize

\bibliographystyle{IEEEtran}
\bibliography{ursi-cs} 

\begin{thebibliography}{10}
\providecommand{\url}[1]{#1}
\csname url@samestyle\endcsname
\providecommand{\newblock}{\relax}
\providecommand{\bibinfo}[2]{#2}
\providecommand{\BIBentrySTDinterwordspacing}{\spaceskip=0pt\relax}
\providecommand{\BIBentryALTinterwordstretchfactor}{4}
\providecommand{\BIBentryALTinterwordspacing}{\spaceskip=\fontdimen2\font plus
\BIBentryALTinterwordstretchfactor\fontdimen3\font minus
  \fontdimen4\font\relax}
\providecommand{\BIBforeignlanguage}[2]{{%
\expandafter\ifx\csname l@#1\endcsname\relax
\typeout{** WARNING: IEEEtran.bst: No hyphenation pattern has been}%
\typeout{** loaded for the language `#1'. Using the pattern for}%
\typeout{** the default language instead.}%
\else
\language=\csname l@#1\endcsname
\fi
#2}}
\providecommand{\BIBdecl}{\relax}
\BIBdecl

\bibitem{Jarvis:20184F}
M.~{Jarvis} and R.~{Taylor et~al.}, ``{The MeerKAT International GHz Tiered
  Extragalactic Exploration (MIGHTEE) Survey},'' \emph{Proceedings of Science},
  vol. MeerKAT Science: On the Pathway to the SKA, Stellenbosch, 25-27 May
  2016, p.~6, 2016.

\bibitem{mightee}
I.~Heywood and M.~Jarvis~et al., ``\BIBforeignlanguage{English}{{MIGHTEE: total
  intensity radio continuum imaging and the COSMOS / XMM-LSS Early Science
  fields}},'' \emph{\BIBforeignlanguage{English}{Monthly Notices of the Royal
  Astronomical Society}}, vol. 509, p. 2150–2168, 2021.

\bibitem{thompsonbook}
G.~{A. Richard Thompson}, {James M. Moran}, \emph{Interferometry and Synthesis
  in Radio Astronomy}.\hskip 1em plus 0.5em minus 0.4em\relax Weinheim:
  WILEY-VCH Verlag GmbH \& Co. KGaA, 2004.

\bibitem{np-hard}
B.~Natarajan, ``Sparse approximate solutions to linear systems,'' \emph{SIAM
  Journal on Computing}, vol.~24, no.~2, pp. 227--234, 1995.

\bibitem{donoho_cs}
D.~L. Donoho, ``Compressed sensing,'' \emph{IEEE Trans. Inf. Theor.}, vol.~52,
  no.~4, pp. 1289--1306, Apr. 2006.

\bibitem{candestao2016}
E.~J. {Candes} and T.~{Tao}, ``Near-optimal signal recovery from random
  projections: Universal encoding strategies?'' \emph{IEEE Transactions on
  Information Theory}, vol.~52, no.~12, pp. 5406--5425, Dec 2006.

\bibitem{frick}
P.~Frick, D.~Sokoloff, R.~Stepanov, and R.~Beck, ``{Wavelet-based Faraday
  rotation measure synthesis},'' \emph{Monthly Notices of the Royal
  Astronomical Society: Letters}, vol. 401, no.~1, pp. L24--L28, 01 2010.

\bibitem{li20102}
F.~{Li}, S.~{Brown}, T.~J. {Cornwell}, and F.~{de Hoog}, ``The application of
  compressive sampling to radio astronomy - ii. faraday rotation measure
  synthesis,'' \emph{A\&A}, vol. 531, p. A126, 2011.

\bibitem{FRM}
M.~A. {Brentjens} and A.~G. {de Bruyn}, ``Faraday rotation measure synthesis,''
  \emph{A\&A}, vol. 441, no.~3, pp. 1217--1228, 2005.

\bibitem{Andrecut_2011}
M.~Andrecut, J.~M. Stil, and A.~R. Taylor, ``Sparse faraday rotation measure
  synthesis,'' \emph{The Astronomical Journal}, vol. 143, no.~2, p.~33, dec
  2011.

\bibitem{akiyama2018faraday}
K.~Akiyama, T.~Akahori, Y.~Miyashita, S.~Ideguchi, R.~Yamaguchi, S.~Ikeda, and
  K.~Takahashi, ``Faraday tomography with sparse modeling,'' 2018.

\bibitem{Ideguchi_2014}
S.~Ideguchi, Y.~Tashiro, T.~Akahori, K.~Takahashi, and D.~Ryu, ``Faraday
  dispersion functions of galaxies,'' \emph{The Astrophysical Journal}, vol.
  792, p.~51, 2014.

\bibitem{10.1093/pasj/psw039}
Y.~Miyashita, S.~Ideguchi, and K.~Takahashi, ``{Performance test of RM CLEAN
  and its evaluation with chi-square value},'' \emph{Publications of the
  Astronomical Society of Japan}, vol.~68, no.~3, 04 2016, 44.

\bibitem{Cooray_2020}
S.~Cooray, T.~T. Takeuchi, T.~Akahori, Y.~Miyashita, S.~Ideguchi, K.~Takahashi,
  and K.~Ichiki, ``An iterative reconstruction algorithm for faraday
  tomography,'' \emph{Monthly Notices of the Royal Astronomical Society}, vol.
  500, no.~4, p. 5129–5141, Nov 2020.

\bibitem{fista}
A.~Beck and M.~Teboulle, ``A fast iterative shrinkage-thresholding algorithm
  for linear inverse problems,'' \emph{SIAM Journal on Imaging Sciences},
  vol.~2, no.~1, pp. 183--202, 2009.

\bibitem{KARL2005183}
W.~Karl, ``3.6 - regularization in image restoration and reconstruction,'' in
  \emph{Handbook of Image and Video Processing (Second Edition)}, A.~BOVIK,
  Ed.\hskip 1em plus 0.5em minus 0.4em\relax Burlington: Academic Press, 2005,
  pp. 183--V.

\bibitem{Hansen00thel-curve}
P.~C. Hansen, ``The l-curve and its use in the numerical treatment of inverse
  problems,'' in \emph{in Computational Inverse Problems in Electrocardiology,
  ed. P. Johnston, Advances in Computational Bioengineering}.\hskip 1em plus
  0.5em minus 0.4em\relax WIT Press, 2000, pp. 119--142.

\bibitem{Belge_2002}
M.~Belge, M.~E. Kilmer, and E.~L. Miller, ``Efficient determination of multiple
  regularization parameters in a generalized l-curve framework,'' \emph{Inverse
  Problems}, vol.~18, no.~4, pp. 1161--1183, jul 2002.

\bibitem{SHI201872}
Y.~Shi, S.~Y. Low, and K.~F. {Cedric Yiu}, ``Hyper-parameterization of sparse
  reconstruction for speech enhancement,'' \emph{Applied Acoustics}, vol. 138,
  pp. 72--79, 2018.

\bibitem{2014MNRAS.439.3591C}
R.~E. {Carrillo}, J.~D. {McEwen}, and Y.~{Wiaux}, ``{PURIFY: a new approach to
  radio-interferometric imaging},'' \emph{Monthly Notices of the Royal
  Astronomical Society}, vol. 439, pp. 3591--3604, Apr. 2014.

\bibitem{purify2}
L.~Pratley, J.~D. McEwen, M.~d'Avezac, R.~E. Carrillo, A.~Onose, and Y.~Wiaux,
  ``{Robust sparse image reconstruction of radio interferometric observations
  with purify},'' \emph{Monthly Notices of the Royal Astronomical Society},
  vol. 473, pp. 1038--1058, 2017.

\bibitem{Lee2019}
G.~R. Lee, R.~Gommers, F.~Waselewski, K.~Wohlfahrt, and A.~O’Leary,
  ``Pywavelets: A python package for wavelet analysis,'' \emph{Journal of Open
  Source Software}, vol.~4, no.~36, p. 1237, 2019.

\bibitem{uwt-denoising}
N.~A. Golilarz and H.~Demirel, ``Image de-noising using un-decimated wavelet
  transform (uwt) with soft thresholding technique,'' in \emph{2017 9th
  International Conference on Computational Intelligence and Communication
  Networks (CICN)}, 2017, pp. 16--19.

\end{thebibliography}

\end{document}